# Copula Mixture Model for Dependency-seeking Clustering

**Mélanie Rey** MELANIE.REY@UNIBAS.CH
**Volker Roth** VOLKER.ROTH@UNIBAS.CH
Department of Mathematics and Computer Science, University of Basel, Basel, Switzerland

## Abstract

We introduce a copula mixture model to perform dependency-seeking clustering when co-occurring samples from different data sources are available. The model takes advantage of the great flexibility offered by the copulas framework to extend mixtures of Canonical Correlation Analysis to multivariate data with arbitrary continuous marginal densities. We formulate our model as a non-parametric Bayesian mixture, while providing efficient MCMC inference. Experiments on synthetic and real data demonstrate that the increased flexibility of the copula mixture significantly improves the clustering and the interpretability of the results.

## 1. Introduction

When different types of measurements concerning a same underlying phenomenon are available, often appearing in the form of co-occurring samples, combining them is more informative than analysing them separately. First, if we assume that these different measurements, also referred to as the different views, are generated by several data sources with independent noise, analysing them jointly can increase the signal to noise ratio. Second, only a combined analysis can take into consideration the dependencies existing between the different types of measurements. As pointed out in Klami & Kaski (2007), possible dependencies between the views often contain some of the most relevant information about the data. Dependency modelling captures what is common between the views, i.e. the shared underlying signal, and in many applications where several experiments are designed to measure the same object this shared aspect is the focus of interest.

Appearing in *Proceedings of the 29$^{th}$ International Conference on Machine Learning*, Edinburgh, Scotland, UK, 2012.

The task of detecting dependencies has traditionally been solved by Canonical Correlation Analysis (CCA). This method can however detect only global linear dependency. When the data express not only one global dependency but different local dependencies, a mixture formulation is more adequate. Fern et al. (2005) introduces a mixture of local CCA model which groups pairs of points expressing together a particular linear dependency between the two views. This model is adapted to cases where the data express several different local correlations, but it still focuses exclusively on linear dependencies since it assumes that within each cluster the two views are linearly correlated.

Dependency-seeking clustering goes one step further in the generalisation process by assuming that the views become independent when conditioned on the cluster structure. The aim is to perform clustering in the joint space of the multiple views, while focussing explicitly on inter-view dependencies [1]. In the case of two views, the objective is then to group the co-occurring pairs of datapoints according to their inter-view dependency pattern such that when the cluster assignments are known these views become independent. As a consequence, the group structure now has a semantic interpretation in terms of dependency with the partition capturing the dependencies.

The starting point of existing dependency-seeking methods is the probabilistic interpretation of CCA given in Bach & Jordan (2005). In Klami & Kaski (2007) a Dirichlet prior Gaussian mixture for dependency-seeking clustering is introduced. However, as pointed out in Klami et al. (2010), when the data are not normally distributed, this method can suffer from a severe model mismatch problem. On application to non-normally distributed data these models have to increase the number of clusters to achieve a reasonable fit. Additional clusters are used to com-pensate for the inadequate Gaussian assumption. The components of these mixtures will not only be used to reflect differences in dependence structures but will also be used to approximate a non-Gaussian distribution. As a result some points expressing a similar inter-view dependence can be assigned to different groups and the interpretation of the clusters in terms of dependencies is lost. Moreover, the model needs to find a compromise between the cluster homogeneity and the approximation of a non-Gaussian mixture, so that non-homogenous clusters might emerge. Figure 1 illustrates how several Gaussian components can be used to approximate a beta density. An exponential family dependency-seeking method is proposed in Klami et al. (2010) to overcome this problem. This model can however be too restrictive when the views are multidimensional. Although the 1-dimensional exponential family covers many interesting distributions, only a few of them have convenient multivariate forms. In particular their dependence structure between dimensions is often very restrictive. Another restriction of that model is that all the dimensions in all the views must have the same univariate distribution whereas in practice different data sources are likely to produce differently distributed data.

[1] The term inter-view dependencies refers to the dependence structure between the different views, whereas intra-view dependencies refers to the dependence structure between the different dimensions of one view.

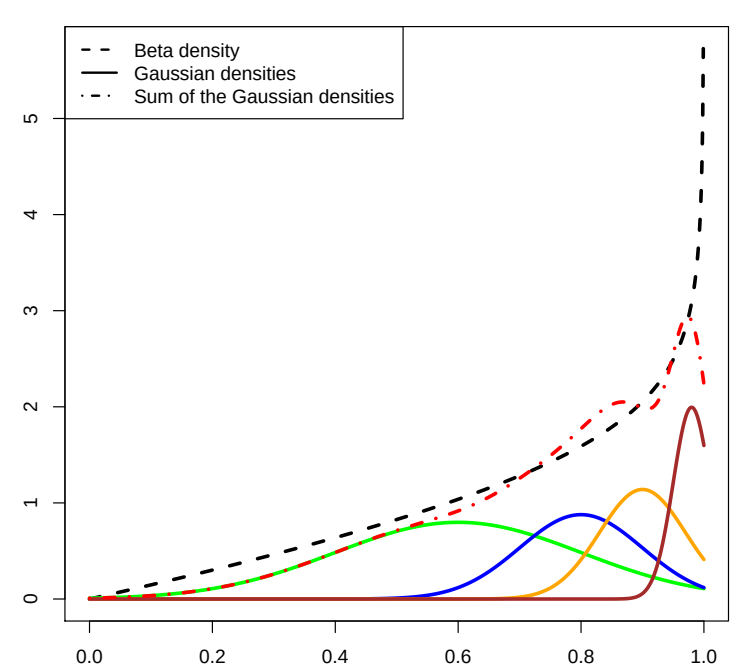

*Figure 1.* Gaussian components approximating a beta density.

To overcome these limitations we take advantage of the copulas framework to build a dependency-seeking clustering method suitable for data with any type of continuous densities. We use Gaussian copulas to construct Dirichlet prior mixtures of multivariate distributions with arbitrary continuous margins, the only restriction being that a density must exist. The model combines the adaptability of Bayesian non-parametric mixtures with the flexibility of copula-based distributions. Our approach focusses on Gaussian copulas for two main reasons. Firstly, their parametrisation using a correlation matrix covers many different dependence patterns ranging from independence to comonotonicity (perfect dependence). Secondly, the model can be reformulated using multivariate Gaussian latent variables which enables efficient MCMC inference.

## 2. Dependency-seeking clustering

Consider a $p$-dimensional random vector (rv) $X$ and a $q$-dimensional rv $Y$ which constitute two different sources of information about an object of interest. For example, several corporal measurements of a patient and the levels of different drugs administrated can serve as two sources of information about a medical treatment. We assume that $X$ and $Y$ have co-occurring samples $(x_1, \ldots, x_n)$ and $(y_1, \ldots, y_n)$ with $x_i \in \mathbb{R}^p$ and $y_i \in \mathbb{R}^q$, $i = 1, \ldots, n$. The probabilistic interpretation of CCA given by Bach & Jordan (2005) uses the following latent variable formulation:

$$Z \sim \mathcal{N}_d\left(0, I_d\right),$$
$$\left(X, Y\right)|Z \sim \mathcal{N}_{p+q}\left(WZ + \mu, \Psi\right),$$

where $\mu = (\mu_x, \mu_y) \in \mathbb{R}^{p+q}$, $W = \begin{pmatrix} W_x \\ W_y \end{pmatrix} \in \mathbb{R}^{(p+q)\times d}$, $1 \leq d \leq \min\left(p, q\right)$ and the covariance matrix $\Psi$ has a block diagonal form:

$$\Psi = \left( \begin{array}{cc} \Psi_x & 0 \\ 0 & \Psi_y \end{array} \right). \tag{1}$$

They showed that the maximum likelihood estimate of $W$ is connected to the canonical directions and correlations:

$$\hat{W}_x = \tilde{\Sigma}_x U_x M_x, \quad \hat{W}_y = \tilde{\Sigma}_y U_y M_y,$$

where $\tilde{\Sigma}_x$, $\tilde{\Sigma}_y$ are the sample covariance matrices, and $U_x$ and $U_y$ are the first $d$ canonical directions. $M_x$ and $M_y$ are matrices such that $M_x M_y^T = P_d$ where $P_d$ is the diagonal matrix containing the first $d$ canonical correlations. Based on the above formulation, the following dependency-seeking clustering model is derived in Klami & Kaski (2008) :

$$Z \sim \text{Mult}\left(\theta\right), \tag{2}$$
$$\left(X, Y\right)|Z \sim \mathcal{N}_{p+q}\left(\mu_z, \Psi_z\right), \tag{3}$$

where $\Psi_z$ has a block structure as in (1):

$$\Psi_z = \left( \begin{array}{cc} \Psi_{zx} & 0 \\ 0 & \Psi_{zy} \end{array} \right). \tag{4}$$

and $\mu_z$ is a mean vector depending on $Z$. The latent variable $Z$ now represents the clustering assignment. A key property of this model is the block diagonal structure of the covariance matrix $\Psi_z$. This special form implies that given the cluster assignment the two views are independent, thereby enforcing the cluster structure to capture all the dependencies. This model however explicitly makes a conditional Gaussian assumption and can perform badly when data within a cluster are non-normally distributed as mentioned in section 1. To relax this normality assumption, we present a dependency-seeking clustering model constructed using Gaussian copulas which can be applied to almost any type of continuous data.

## 3. Copulas and Gaussian copulas

A multivariate distribution is constituted of univariate random variables related to each other by a dependence mechanism. Copulas provide a framework to separate the dependence structure from the marginal distributions. Formally, a $d$-dimensional copula is a multivariate distribution function $C : [0,1]^d \to [0,1]$ with standard uniform margins. The following theorem Sklar (1959) states the relationship between copulas and multivariate distributions.

**Theorem 1.** (Sklar) *Let $F$ be a joint distribution function with margins $F^1, \ldots, F^d$. Then there exists a copula $C : [0,1]^d \to [0,1]$ such that*

$$F\left(x^1, \ldots, x^d\right) = C\left(F^1\left(x^1\right), \ldots, F^d\left(x^d\right)\right). \quad (5)$$

*Moreover, if the margins are continuous, then this copula is unique. Conversely, if $C$ is a copula and $F^1, \ldots, F^d$ are univariate distribution functions, then $F$ defined as in (5) is a multivariate distribution function with margins $F^1, \ldots, F^d$.*

Gaussian copulas constitute an important class of copulas. If $F$ is a Gaussian distribution $\mathcal{N}_d(\mu, \Sigma)$ then the corresponding $C$ fulfilling equation (5) is a Gaussian copula. Since Gaussian copulas are invariant to strictly increasing transformations, the copula of $\mathcal{N}_d(\mu, \Sigma)$ is the same as the copula of $\mathcal{N}_d(0, P)$ as mentioned in McNeil et al. (2005), where $P$ is the correlation matrix corresponding to the covariance matrix $\Sigma$. Thus a Gaussian copula is uniquely determined by a correlation matrix $P$ and we denote a Gaussian copula by $C_P$. Using theorem 1 with $C_P$, we can construct multivariate distributions with arbitrary margins and a Gaussian dependence structure. These distributions, called meta-Gaussian distributions, provide a natural way to extend models based on a multivariate normality assumption.

When using a Gaussian copula we do not attempt to directly model the correlation of the original variables, but instead we first apply the transformation $\Phi^{-1}\left(F^j(\,.\,)\right)$ to every margin to obtain normally distributed variables $\Phi^{-1}\left(F^j(X^j)\right) \sim \mathcal{N}_1(0,1)$, where $\Phi$ is the standard Gaussian cumulative distribution function, and then use $P$ to describe their correlation. We finally note that zero values in $P$ encode independence between the corresponding marginal variables. Therefore, if $P$ has a block diagonal structure as in (1), the conditional independence of $X|Z$ and $Y|Z$, which was a key property of equation (3), will be preserved in a meta-Gaussian model.

Multivariate distributions constructed using Theorem 1 do not necessarily possess a density function. When a density exist it can be written as:

$$f(x^1, \ldots, x^d) = c\left(F^1(x^1), \ldots, F^d(x^d)\right) \prod_{j=1}^{d} f^j(x^j), \quad (6)$$

where

$$c(u^1, \ldots, u^d) = \frac{\partial C(u^1, \ldots, u^d)}{\partial u^1 \ldots \partial u^d}, \quad (7)$$

is the copula density of $C$. For cases where $c$ has a simple closed form we can obtain an analytical expression for $f$ using (6). This is true for the multivariate normal case and equation (6) becomes:

$$f(x) = |P|^{-\frac{1}{2}} \exp\left\{-\frac{1}{2}\tilde{x}^T(P^{-1} - I)\tilde{x}\right\} \prod_{j=1}^{d} f^j(x^j), \quad (8)$$

where $\tilde{x}^j = \Phi^{-1}\left(F^j(x^j)\right)$, $x = (x^1, \ldots, x^d)$, $\tilde{x} = (\tilde{x}^1, \ldots, \tilde{x}^d)$. We denote this density by $\mathcal{M}(\theta, P)$, where $\theta$ is the vector containing all parameters of the marginal distributions.

## 4. Multi-view clustering with meta-Gaussian distributions

### 4.1. Model specification

Consider the two rv $X = \left(X^1, \ldots, X^p\right)$ and $Y = \left(Y^1, \ldots, Y^q\right)$. We assume their joint distribution is a Dirichlet prior mixture (DPM) given by:

$$f_{(X,Y)}(x,y) = \int\int f_{(X,Y)|\theta,P}(x,y) \mathrm{d}\mu_{\theta,P} \mathrm{d}\mu_G(\lambda, G_0),$$

where $\mu_G$ is the distribution of a Dirichlet process (Ferguson, 1973) with base distribution $G_0$ and concentration parameter $\lambda$. The novelty here is the choice of $f_{(X,Y)|\theta,P}$. We model the marginal distributions and the dependence structure separately to allow for more freedom:

1. The margins can be arbitrary continuous distributions (providing the corresponding density exists):

$$X^j|\theta = X^j|\theta_x^j \sim F^j_{X|\theta},\ j = 1, \dots, p,$$
$$Y^j|\theta = Y^j|\theta_y^j \sim F^j_{Y|\theta},\ j = 1, \dots, q,$$

   where $\theta = \left(\theta_x^1, \dots, \theta_x^p, \theta_y^1, \dots, \theta_y^q\right)$. Note here that $F^j_{X|\theta}$ can be different types of distributions for the multiple dimensions $j$.

2. The dependence structure is then specified by a Gaussian copula $C_P$ with correlation matrix $P$ having a block diagonal structure as in (1).

3. Finally the constructed multivariate distribution will have the form:

$$F_{(X,Y)|\theta,P}(x,y) = C_P\left(F^1_{X|\theta}\left(x^1\right), \dots, F^q_{Y|\theta}\left(x^q\right)\right). \tag{9}$$

### 4.2. Bayesian inference

Separating the modelling task between specification of the margins and specification of the dependence structure simplifies the choice of the prior distributions. If we assume *a priori* independence for $\theta$ and $P$ we can specify prior distributions for the margins and separately choose a prior for the parameters of the copula $C_P$. We specify independent prior distributions for the blocks $P_x$ and $P_y$, where $P = \begin{pmatrix} P_x & 0 \\ 0 & P_y \end{pmatrix}$. For $P_x$ and $P_y$ we choose the marginally uniform prior given in Barnard et al. (2000). This prior is a multivariate distribution on the space of correlation matrices with uniform margins, i.e. $P_{ij}$ is a uniform variable for $i \neq j$, and is connected to the inverse-Wishart distribution: if a covariance matrix $\Psi \in \mathbb{R}^{d\times d}$ is standard inverse-Wishart distributed with parameter $I_d$ and $d+1$ degrees of freedom, then the corresponding correlation matrix $R$ follows the marginally uniform prior distribution.

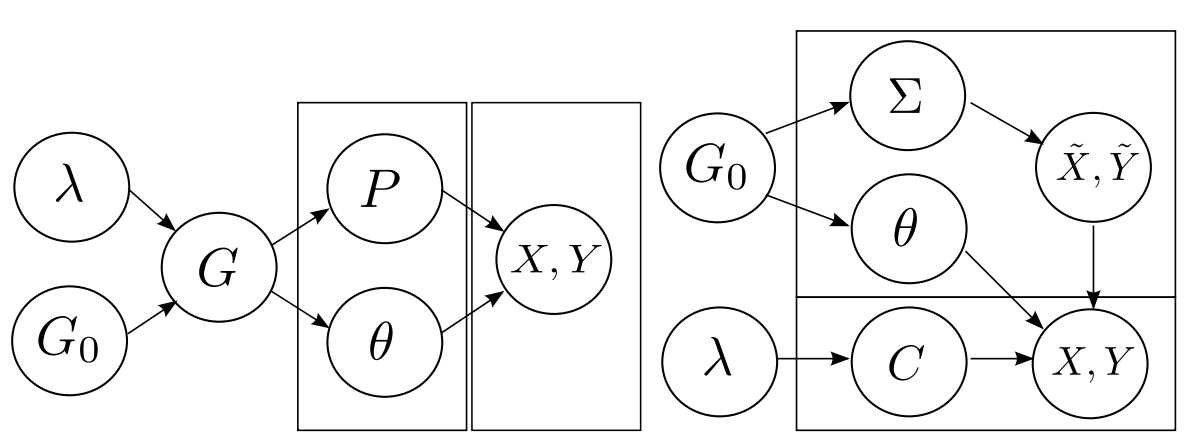

*Figure 2.* Graphical representation of the infinite copula mixture model with base measure $G_0$ and concentration $\lambda$. Left side: the original model, right side: the model augmented for sampling, where $C$ denotes cluster assignment.

Inference can be done using MCMC sampling methods for Dirichlet process mixture models. We use a sampling scheme for models with non-conjugate prior given in Neal (2011). The method, detailed in Algorithm 1, is composed of three steps: a modified Metropolis-Hastings step, partial Gibbs sampling updates and an update of the parameters $\theta, P$. In the third step we need to update the parameters of every cluster according to their posterior distribution. Since we cannot sample directly from this conditional posterior we developed a sampling scheme similar to the algorithm proposed in Hoff (2007). The main idea is to overparametrize the model by introducing a normally distributed latent vector $(\tilde{X}, \tilde{Y})$. The variables in the complete model are then given by:

$$(\tilde{X}, \tilde{Y}) \sim \mathcal{N}_{p+q}(0, \Sigma),$$
$$(X, Y) \sim \mathcal{M}(\theta, P),$$
$$C \sim \mathrm{CRP}(\lambda),$$

where $\Sigma$ is a covariance matrix with corresponding correlation matrix $P$ and $C$ denotes the cluster assignments following a Chinese restaurant process distribution. Figure 2 gives a representation of the complete model. In the MCMC scheme we can easily sample $\Sigma$ conditioned on $(X, Y)$, $(\tilde{X}, \tilde{Y})$ and $\theta$, since we can use the conjugacy property of prior and conditional likelihood. A sample of the correlation matrix can be otained as $\mathcal{P}(\tilde{X})$, the correlation matrix of the random vector $\tilde{X}$. The posterior updates of the parameters are detailed in Algorithm 2. The notations $\theta^{\star j}, \mathcal{P}(\tilde{X})^{\star j}$ are used to emphasize that the corresponding vector or matrix is considered as a function of $\theta^j, \tilde{X}^j$ and parameters for the other dimensions are treated as constants.

## 5. Experiments

### 5.1. Simulated data

We simulate two different 2-dimensional multi-view data sets with Gaussian intra-view dependence structure. The marginal distributions are Gaussian in the first view, and beta or exponential in the second. Each data set is composed of two clusters which can be identified only by considering the inter-view dependencies. We first simulated data points with a single cluster structure in each view but a strong positive dependence between the first dimensions of the views, i.e. between $X^1$ and $Y^1$. In a second step we separated the data in two groups of unequal size and randomly permuted their order within groups to suppress any inter-view dependency within these groups. Figure 3 (bottom left panel) shows the resulting cluster structure in the joint space of the two views recovered by the copula mixture model. Parameters used for the simulations can be found in Table 1.

**Algorithm 1** Markov Chain Sampling

$C_1, \ldots, C_n$ are the latent variables of the cluster assignments.
$\theta^{C_i}$ and $P^{C_i}$ are the parameters for cluster $C_i$.
$n_{-i,c}$ is the number of datapoints in cluster $c$ excluding observation $i$.
$C_{-i} = \{C_1, \ldots, C_{i-1}, C_{i+1}, \ldots, C_n\}$.
**repeat**
  **for** $i = 1, \ldots, n$ **do**
    **if** there exists $k$ such that $C_k = C_i$ **then**
      Create a new cluster $C_i^*$ with parameters $\theta^*$ and $P^*$ drawn from $G_0$;
      Change $C_i$ to $C_i^*$ with probability $\min\left(1, \frac{\lambda}{n-1} \frac{f_{(X,Y)|\theta^*,P^*}(x,y)}{f_{(X,Y)|\theta^{C_i},P^{C_i}}(x,y)}\right)$;
    **else**
      Draw $C_i^*$ from $C_{-i}$ with $\mathrm{P}(C_i^* = c) = n_{-i,c}/(n-1)$. Change $C_i$ to $C_i^*$ with probability $\min\left(1, \frac{n-1}{\lambda} \frac{f_{(X,Y)|\theta^*,P^*}(x,y)}{f_{(X,Y)|\theta^{C_i},P^{C_i}}(x,y)}\right)$;
    **end if**
  **end for**
  **for** $i = 1, \ldots, n$ **do**
    **if** there exists $k$ such that $C_k = C_i$ **then**
      Choose a new value for $C_i$ with $\mathrm{P}(C_i^* = c) \propto \frac{n_{-i,c}}{(n-1)} f_{(X,Y)|\theta^c,P^c}(x,y)$;
    **end if**
  **end for**
  **for** $c \in \{C_1, \ldots, C_n\}$ **do**
    Update the parameters $\theta^c$ and $P^c$ as described in Algorithm 2.
  **end for**
**until** stopping criterion

**Algorithm 2** Posterior updates of $(\theta, P) \mid (X, Y)$

For clarity we omit the cluster index $c$.
1.*Sample* $\theta|\Sigma, (\tilde{X}, \tilde{Y}), (X, Y)$
**for** $j = 1, \ldots, p$ **do**
  Draw $\theta^j$ using Metropolis-Hastings;
  $\theta^j \sim f(\theta^j|\theta^{-j}, \tilde{X}, X) \propto \mathcal{M}(\theta^{\star j}, \mathcal{P}(\tilde{X}))\pi(\theta^j)$
**end for**
Apply the same procedure for $Y$;
2. *Sample* $(\tilde{X}, \tilde{Y})|\theta, \Sigma, (X, Y)$
**for** $j = 1, \ldots, p$ **do**
  Draw $\tilde{X}^j$ using Metropolis-Hastings;
  $\tilde{X}^j \sim f(\tilde{X}^j|\tilde{X}^{-j}, \theta, \Sigma, X) \propto \mathcal{M}(\theta, \mathcal{P}(\tilde{X})^{\star j})\mathcal{N}(0, \Sigma)$
**end for**
Apply the same procedure for $\tilde{Y}$;
3. *Sample* $\Sigma|(\tilde{X}, \tilde{Y}),\ \theta, (X, Y)$:
Draw $\Sigma_x \sim \mathcal{N}(0, \Sigma_x)\mathcal{IW}(p+1, I_p)$
$\sim \mathcal{IW}(I_p + \sum_{i=1}^n \tilde{X}_i \tilde{X}_i^T, p+1+n)$;
Apply the same procedure to obtain $\Sigma_y$.

*Table 1.* Parameters used for the simulations.

| | | | |
|---|---|---|---|
| Simulation 1 | view 1: Normal | $\mu$ | $(0,0)$ |
| | | $\sigma^2$ | $(1,1)$ |
| | | $(P_x)_{12}$ | 0.9 |
| | view 2: Beta | $\alpha$ | $(3,1)$ |
| | | $\beta$ | $(1,10)$ |
| | | $(P_y)_{12}$ | $-0.5$ |
| Simulation 2 | view 1: Normal | $\mu$ | $(0,0)$ |
| | | $\sigma^2$ | $(1,1)$ |
| | | $(P_x)_{12}$ | 0.9 |
| | view 2: Exponential | $\lambda$ | $(2.5, 2.5)$ |
| | | $(P_y)_{12}$ | 0.9 |

We compared the copula mixture (CM) with three other methods: a Dirichlet prior Gaussian mixture for dependency-seeking clustering (GM) as derived in Klami & Kaski (2007), a non-Bayesian mixture of canonical correlation models (CCM) (Vrac, 2010) (Fern et al., 2005) and a variational Bayesian mixture of robust CCA models (RCCA) (Viinikanoja et al., 2010). CCM and RCCA both assume that the number of clusters is known or can be determined as explained in (Viinikanoja et al., 2010). In our comparison experiments we gave as input for both methods the correct number of clusters, giving them the advantage of this extra knowledge. Results presented in Figure 4 show that CM applied with the correct marginal distributions' form produces a better classification. GM does not perform well on those data sets because the number of clusters is overestimated; the model compensates for the inadequate Gaussian assumption by multiplying the number of components and additional clusters are created to approximate non-Gaussian distributions. Since the number of clusters in a Dirichlet prior Gaussian mixture can be reduced by imposing a too-strong prior on the variances, we modified the prior information to enforce artificially high variances in the second view until the mixture is forced to create no more than two clusters. We report both results obtained with less (GM1) and more (GM2) informative priors. As can be seen in Figure 3, when strong prior information is used to artificially reduce the number of clusters, the GM cannot recover the true cluster structure. CCM and RCCA used with the correct number of clusters as input perform comparatively, or better than the GM but clearly worse than CM for those data sets having non-linear inter-view dependencies.

### 5.2. Real data

We perform a combined analysis of two data sets providing information about the regulation of gene expression in yeast under heat shock; each data set being treated as one view. The first data set (published in Gasch et al. (2000)) provides genes expression values measured at 4 time points. The second data set (given in Harbison et al. (2004)) contains binding affinity scores for interactions between these genes and 6 different binding factors. Similar data have already been analysed in Klami & Kaski (2007). 5360 genes present in both views are clustered using a Gaussian dependency-seeking clustering model (GM) and using the copula mixture (CM). CM uses Gaussian marginals in the first view and beta marginals in the second view. Here the choice of the beta distribution is motivated by the fact that observations in the second view are restricted to the $[0, 1]$ interval. For the univariate Gaussian margins we choose normal and inverse-gamma priors for mean and variance respectively, whereas for the beta margins both shape parameters have gamma priors. GM uses the standard conjugate prior [2].

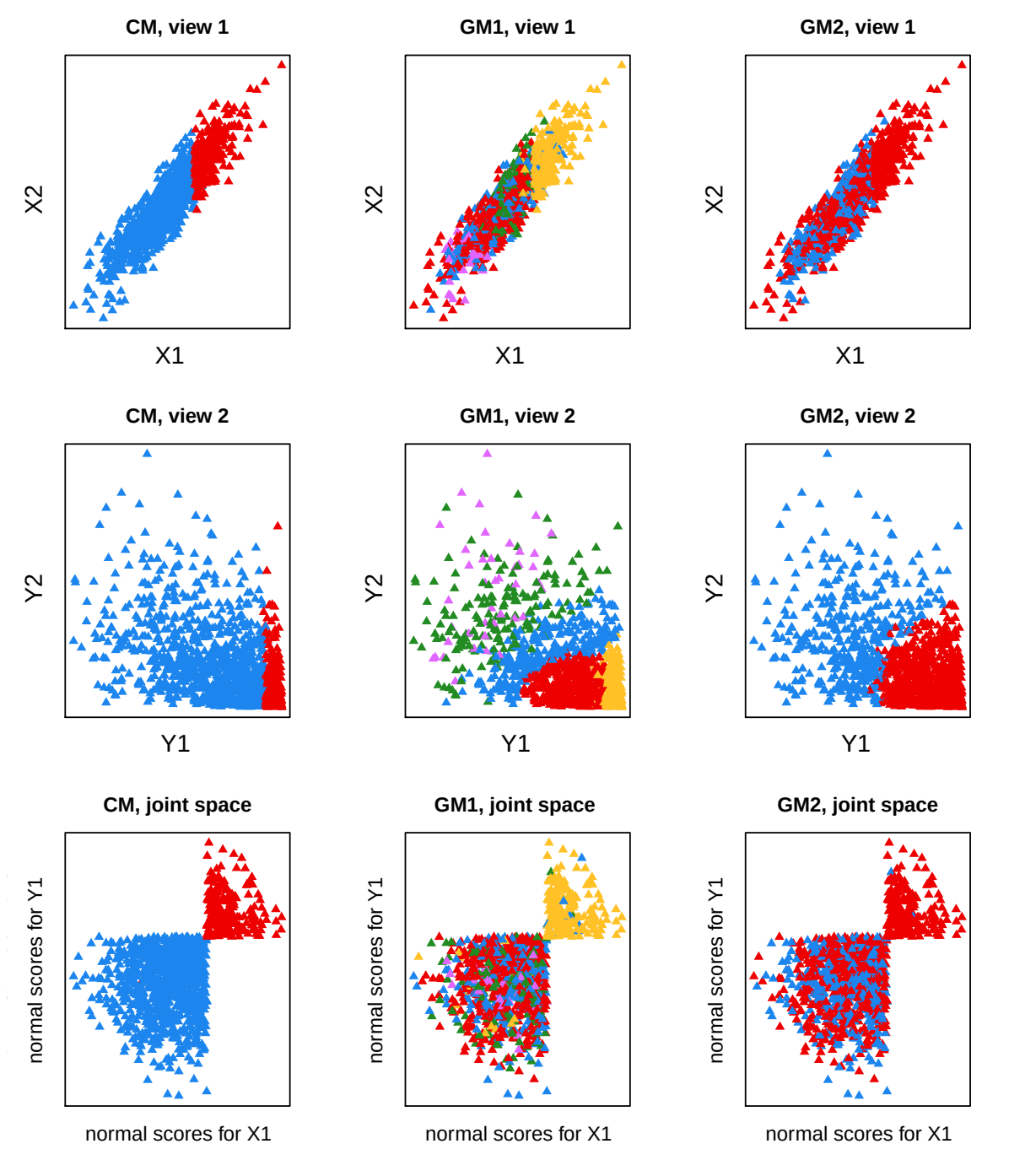

*Figure 3.* Scatterplot of the simulated data in the Gaussian view (first view, top panel), in the beta view (second view, middle panel) and in the joint space of the normal scores for the two views where the two clusters can be clearly identified (bottom panel). The clustering results are shown for the copula mixture (CM) and the Gaussian mixture with two different priors (GM1 and GM2). CM perfectly recovers the true cluster structure, whereas a model mismatch problem prevents GM to find the correct clustering.

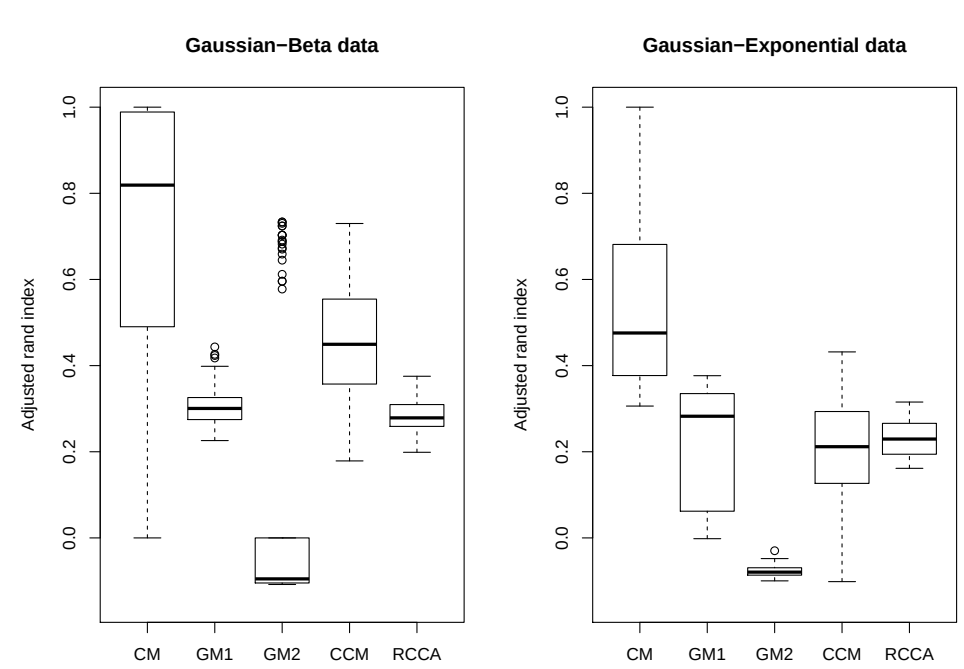

*Figure 4.* Boxplot of the adjusted rand index over 100 (Gaussian-beta data on the left panel) and 50 (Gaussian-exponential data on the right panel) simulations for the copula mixture (CM), the Gaussian mixture with two different priors (CM1 and CM2), the non-Bayesian mixture of CCA (CCM), and the robust CCA mixture (RCCA). Friedman's test with post-hoc analysis rejected, for both experiments, the null hypothesis of equal medians between CM and every other method (P-value $< 0.005$).

For different values of the concentration parameter $\lambda \in \{0.01, 0.1, 1, 5, 10\}$, CM consistently estimates 8 clusters whereas GM estimated between 13 and 15 clusters. In this section we report the results obtained with $\lambda = 1$. As we observed with the simulated data more clusters need to be created by the Gaussian mixture to compensate for the model mismatch. This phenomenon is illustrated in Figure 5. The interpretation of the clustering then becomes very arduous since these additional clusters cannot be distinguished from those capturing the dependencies. Another interpretation problem clearly arises in the Gaussian model when we look at the estimated intra-view correlations. Two negative effects accumulate here; first correlation can be an inadequate dependence measure for non-normally distributed data, and second the additional split in many components can change the cluster-specific intra-view dependence as illustrated in Figure 6.

To understand what information one could gain by dependency-clustering, we perform three additional clustering of the same data: first we cluster the data-points on each view separately, then we cluster them in the complete product space of the joint views, i.e. without imposing the constraint of a block structure on the correlation matrix. Priors and hyperparameters are kept constant over experiments. CM finds four clusters in the first view as well as in the second view. Clustering in the product space with full correlation matrix again leads to four groups. Figure 7 illustrates how the three main clusters found in the complete product space are further separated by dependency-seeking clustering, showing dependencies between the two views.

[2] The use of conjugate prior does not, in general, increase the number of clusters as shown in Rasmussen & Görür (2010).

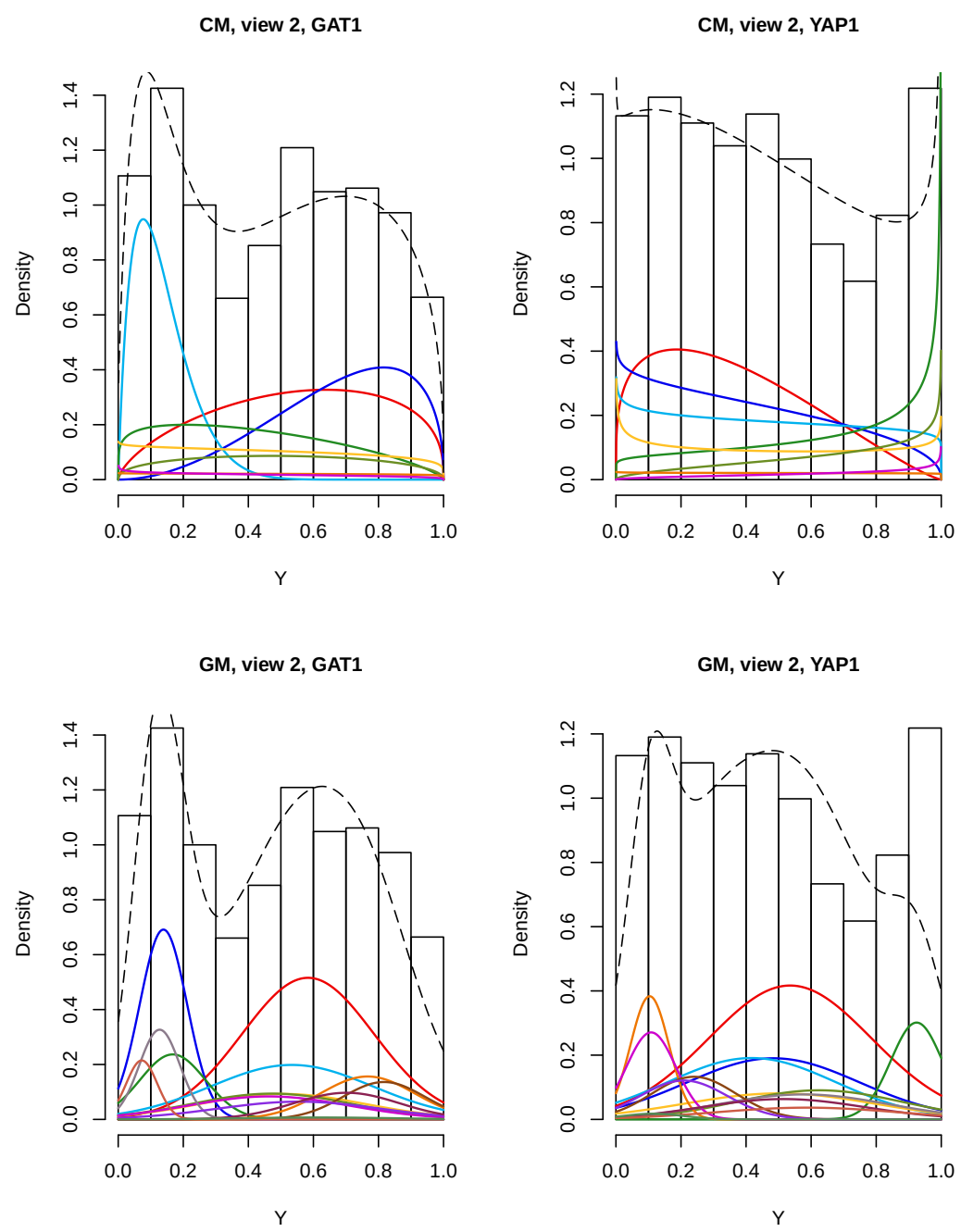

*Figure 5.* Histogram of the binding affinity scores for the binding factors GAT1 and YAP1. The estimated densities of the 8 clusters discovered by CM are represented as colored lines in the top panel. Estimated densities of the 14 clusters found by GM are shown in the bottom panel. The black dashed lines represent the total density resulting of the mixture.

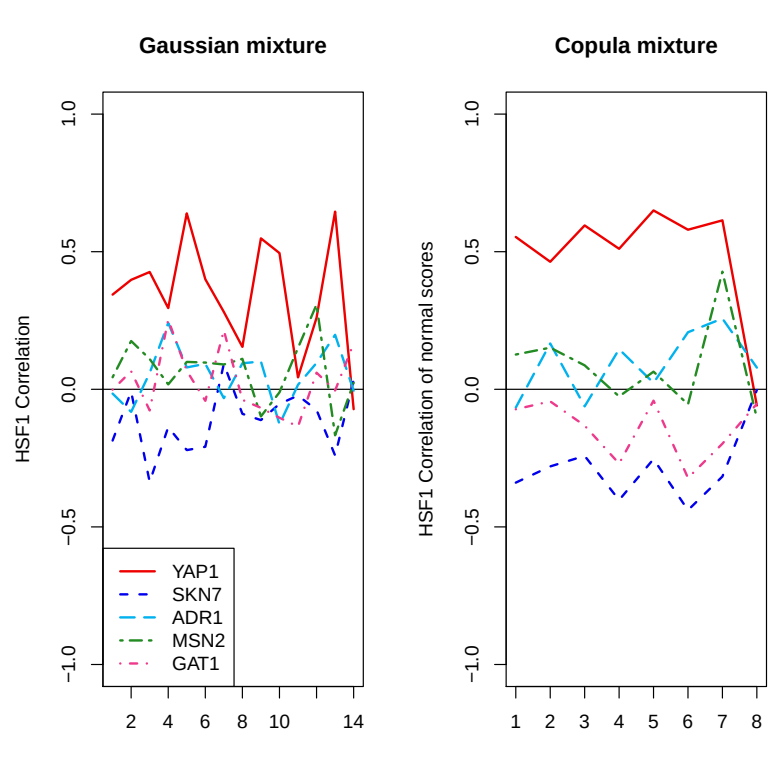

*Figure 6.* Correlations estimated with GM (left panel) and correlations of the normal scores estimated by CM (right panel) between HSF1 and the five other binding factors. In the Gaussian model the correlation between HSF1 and YAP1 seems to vary drastically with the clusters. In CM this correlation has stable positive values for all clusters with the exception of the last cluster. Since the binding factors HSF1 and YAP1 are both activated by the substance *menadione* as explained in Hohmann & Mager (2003), we can expect that their binding affinities are positively correlated and independent of the cluster.

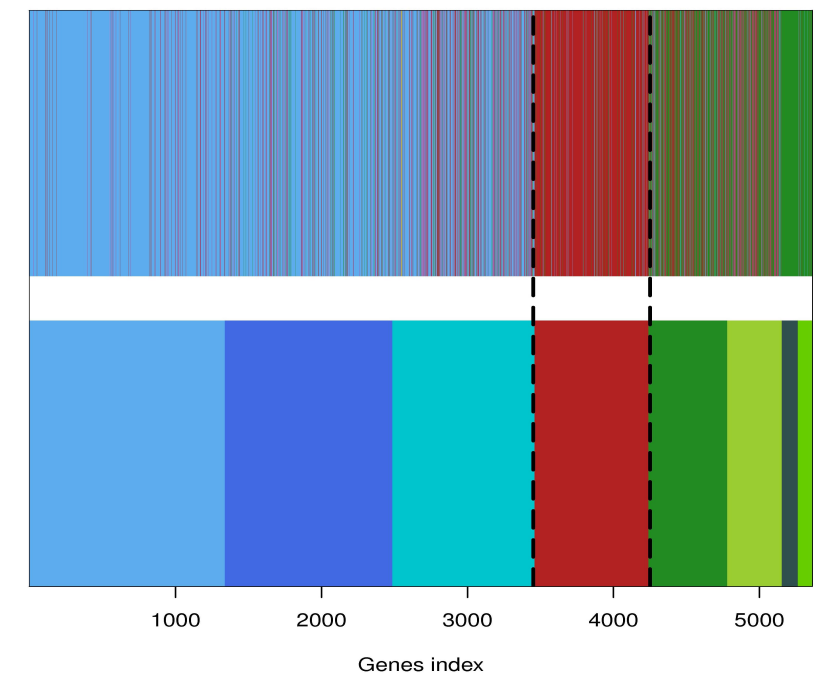

*Figure 7.* The bottom panel represents in different colors the cluster indices for all genes (reordered by cluster assignment) as obtained using dependency-seeking clustering with CM. The top panel shows the cluster indices obtained when clustering in the complete product space, i.e using CM with a full correlation matrix instead of a restricted block diagonal matrix. This illustrates how existing groups are further separated into smaller clusters expressing inter-view dependencies.

As mentioned in section 1, GM cannot exclusively focus on compact clusters because it needs to find a compromise between the cluster homogeneity and the approximation of a non-Gaussian mixture. As a result, non-homogenous clusters might emerge which are needed to fit the margins despite model mismatch. To test if this phenomenon is present here, we perform a gene ontology enrichment analysis (GOEA) using GOrilla (Eden et al., 2009). GOEA is used to test if some of the biological processes associated with the genes are over-represented in the clusters, thereby providing a quality measure for the clustering. The analysis shows that 3 out of 14 clusters (these 3 clusters representing together 17,3% of the data points) found by GM do not express any significant enrichment. By contrast, all 8 clusters produced by CM express a highly significant enrichment and every cluster can be associated with a specific biological processes, e.g. the two largest clusters can be interpreted as groups of genes involved in organelle organization and meiosis respectively. The clear difference in the enrichment analysis results between GM and CM demonstrates that the quality of the clustering is indeed impaired when a model with inadequate margins is used.

## 6. Conclusion

A fundamental aspect in dependency-seeking clustering is that the partition possesses a semantic interpretation in terms of dependency: the dependencies are captured by the cluster structure. This interpretation is however only valid when the model is rich enough to properly fit each view, which can be particularly difficult to achieve for non-Gaussian data with existing models. This task becomes even more arduous when the dimensions of the views increase since the model then needs to adequately fit every margin while allowing for a sufficiently rich intra-view dependence structure. The copula mixture model offers enough flexibility to cover both aspects: the margins can be specified separately for each dimension and the Gaussian copula allows for a wide range of intra-view dependencies. Using a Gaussian copula also facilitates the inference and we provide an efficient MCMC scheme. Experiments on simulated data show that the copula mixture model significantly improves the clustering results. In a large-scale real-world clustering problem of genes expression data and genes binding affinities, the dependency-seeking copula mixture model produces a clustering solution that significantly differs from those obtained on the single views or on the product space, and from that obtained by the standard Gaussian model which clearly suffered from model-mismatch problems. Detailed analysis of the functional annotation of the genes in the clusters discovered by dependency-seeking CM shows that the induced cluster structure allows a plausible biological interpretation in that the groups are clearly enriched by genes involved in distinct biological processes.